\documentclass[prd,twocolumn,showpacs,showkeys,preprintnumbers,floatfix,nofootinbib,superscriptaddress,aps,10pt,reprint]{revtex4-1}

\usepackage{amsmath, graphicx}
\usepackage[colorinlistoftodos]{todonotes}
\usepackage[colorlinks=true, allcolors=blue]{hyperref}
\usepackage{listings, color}

\graphicspath{{figs/}}

\begin{document}
\title{Machine learning estimators for lattice QCD observables}
\author{Boram Yoon}
\email{boram@lanl.gov}
\affiliation{Computer, Computational, and Statistical Sciences Division
CCS-7, Los Alamos National Laboratory, Los Alamos, New Mexico 87545}
\author{Tanmoy Bhattacharya}
\email{tanmoy@lanl.gov}
\affiliation{Theoretical Division
T-2, Los Alamos National Laboratory, Los Alamos, New Mexico 87545}
\author{Rajan Gupta}
\email{rg@lanl.gov}
\affiliation{Theoretical Division 
T-2, Los Alamos National Laboratory, Los Alamos, New Mexico 87545}
\preprint{LA-UR-18-26411}
\begin{abstract}
A novel technique using machine learning (ML) to reduce the
computational cost of evaluating lattice quantum chromodynamics (QCD)
observables is presented. The ML is trained on a subset of background
gauge field configurations, called the labeled set, to predict an
observable $O$ from the values of correlated, but less
compute-intensive, observables $\mathbf{X}$ calculated on the full
sample. By using a second subset, also part of the labeled set, we
estimate the bias in the result predicted by the trained ML
algorithm. 
A reduction in the computational cost by about 7\%--38\% is
demonstrated for two different lattice QCD calculations using the
Boosted decision tree regression ML algorithm: (1) prediction of
the nucleon three-point correlation functions that yield isovector
charges from the two-point correlation functions, and (2) prediction
of the phase acquired by the neutron mass when a small Charge-Parity
(CP) violating interaction, the quark chromoelectric dipole moment
interaction, is added to QCD, again from the two-point correlation
functions calculated without CP violation.

\end{abstract}
\maketitle

\section{Introduction}
\label{sec:intro}

Simulations of lattice QCD provide values of physical observables from
correlation functions calculated as averages over gauge field
configurations, which are generated using a Markov chain Monte Carlo
method using the action as the Boltzmann weight~\cite{Wilson:1974sk,
  Creutz:1980zw}.  Each measurement is computationally expensive and a
standard technique to reduce the cost is to replace the ``high
precision'' average of an observable \(O\) by a ``low precision''
(LP) version of it, \(O_{\rm LP}\)~\cite{Bali:2009hu,Blum:2012uh}, and
then perform bias correction (BC), i.e., \(\langle O\rangle = \langle
O_{\rm LP}\rangle + \langle O - O_{\rm LP}\rangle\). The method works
because the second term can be estimated with sufficient precision
from a smaller number of measurements if the covariance between \(O\)
and \(O_{\rm LP}\) is positive and comparable to the variance of
\(O\), which is the case if, for example, the fluctuations in both are
controlled by effects common to both. One can replace \(O_{\rm LP}\)
in the above formulation with any quantity; however, improved results
are obtained when a quantity with statistical fluctuations similar to
that of \(O\) is chosen for \(O_{\rm LP}\). Since most underlying
gauge dynamics affect a plethora of observables in a similar way, such
quantities surely exist; the trick, however, is to find suitable sets
of quantities.

Machine learning algorithms (ML) build predictive models from data. In
contrast to conventional curve-fitting techniques, ML does not use a
``few parameter functional family'' of forms for the
prediction. Instead, it searches over the space of functions
approximated using a general form with a large number of free
parameters that require a correspondingly large amount of training
data to avoid overfitting. ML has been successful for various
applications where such data are available, including exotic particle
searches~\cite{Baldi:2014kfa} and $\textrm{Higgs}\rightarrow \tau
\tau$ analyses~\cite{Baldi:2014pta} at the Large Hadron Collider. It
has recently been applied to lattice QCD
studies~\cite{Alexandru:2017czx,Wetzel:2017ooo,Shanahan:2018vcv}.
Here we introduce a general ML method for estimating observables
calculated using expensive Markov chain Monte Carlo simulations of
lattice QCD that reduce the computational
cost.\looseness-1

Consider $M$ samples of independent measurements
of a set of observables $\mathbf{X}_i = \{o_i^1, o_i^2, o_i^3,
\ldots\}$, ${i=1,\ldots,M}$, but the target observable $O_i$
is available only on $N$ of these.
These $N$ are called the \emph{labeled data}, and the remaining $M-N$ are 
called the \emph{unlabeled data} ($UD$). Our goal is
to build a ML model $F$ that predicts the target observable
${O_i\approx O_i^{\textrm{P}}\equiv F(\mathbf{X}_i)}$ by
training a ML algorithm on a subset $N_t<N$ of the labeled
data. The bias-corrected estimate \(\overline
O\) of \(\langle O\rangle\) is then obtained as\looseness-1
\begin{align}
\overline{O} =  \frac{1}{M-N} \sum_{i\in \{UD\}} O^\textrm{P}_i 
+ \frac{1}{N_b} \sum_{i \in \{BC\}} (O_i - O^\textrm{P}_i)\,,
\label{eq:o_unbiased}
\end{align}
where the second sum is over the ${N_{b}\equiv N-N_t}$ remaining
labeled samples that corrects for possible bias. Here
\(O^\textrm{P}_i\) depends explicitly on \(\mathbf{X}_i\) and
implicitly on \(N_t\) and all training data \(\{O_j,
\mathbf{X}_j\}\). For fixed ML model $F$, the sampling variance of
$\overline{O}$ is then given by
\begin{align}
    \sigma_{\overline O}^2 = \frac{\sigma_O^2}N \left\{s^2\frac N{M-N}+ \frac1f [(1-s)^2 + 2s(1-r)]\right\}\,,
\label{eq:var}
\end{align}
where \(\sigma_O^2\) is the variance of \(O_i\),
\(s\equiv\sigma_{O^P}/\sigma_O\) is the ratio of the standard
deviations of the predictor variable \(O^P\) to the true observable
\(O\), \(r\) is the correlation coefficient between these two, and
\(f\equiv N_b/N\) is the fraction of observations held out for bias
correction.  Eq.~\eqref{eq:var} shows that when \(s\approx1\approx
r\), this procedure increases the effective sample size from \(N\),
where \(O_i\) are available, to about \(M-N\). For simplicity, in
deriving Eq.~\eqref{eq:var} we have ignored details such as the
statistical independence of the data.
In this work, we account for
the full error, including the sampling variance of the training and 
the bias correction datasets, by using a bootstrap
procedure~\cite{10.2307/2958830} that independently selects \(N\)
labeled and \(M-N\) unlabeled items for each bootstrap sample.

Two additional remarks regarding bias correction are in order. First, while the
bias correction removes the systematic shift in the prediction, it can
increase the final error; i.e., the systematic error can get converted
to a statistical error. In practice, for the two examples discussed
below, the BC does not increase the error significantly.  Second,
there are two ways of bootstrapping the training and BC samples: (i)
first partitioning the labeled data into training and BC sets and
bootstrapping these and (ii) bootstrapping over the full labeled set
and then partitioning the bootstrap sample. We used the latter approach.

\section{Experiment A: Nucleon Isovector Charges}
\label{sec:EXPa}

\begin{figure}[tbp]
\includegraphics[width=0.48\textwidth]{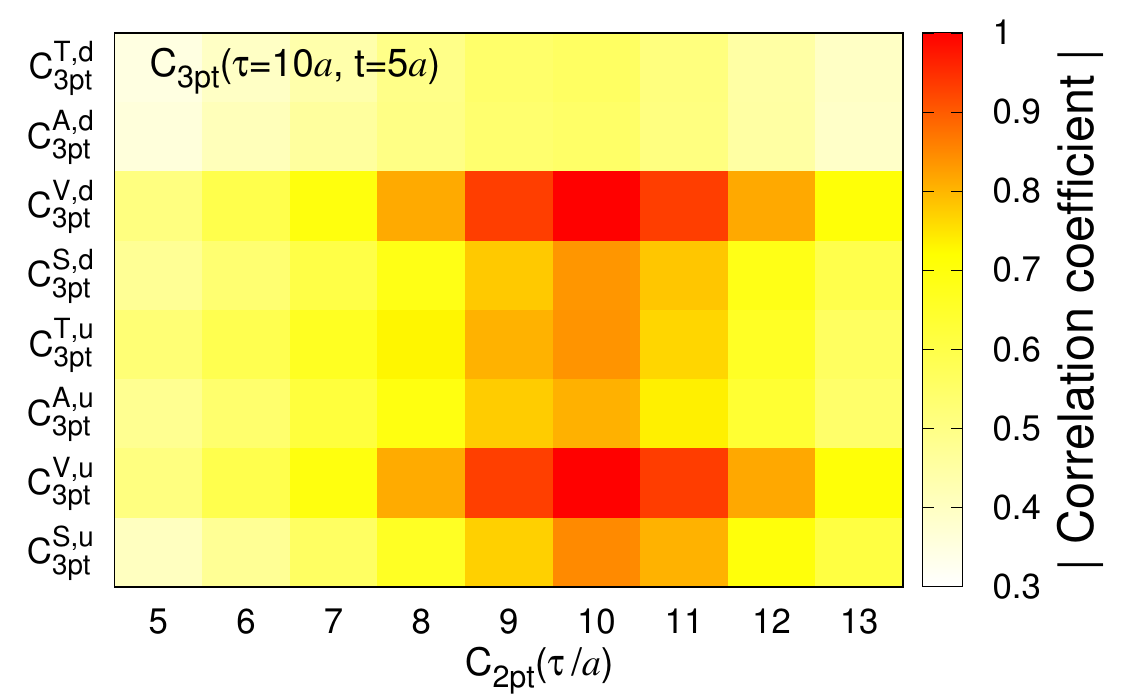}
\caption{Absolute value of the correlation coefficients between the
  proton $C_\textrm{3pt}(\tau=10a,t=5a)$ and the $C_{\rm 2pt}$ at
  various sink time slices $\tau/a=5,6,\ldots,13$.  The data are for the $a09m310$
  ensemble.  The source points of $C_{\rm 2pt}$ and $C_{\rm 3pt}$ are
  fixed at $t=0$. The operator insertion in $C_{\rm 3pt}$ is at $t=5a$
  and the sink is at $\tau=10a$.   
\label{fig:corr_map}}
\end{figure}


For a first example, we demonstrate that this method reduces the
computing cost for the isovector ($u-d$) combination of the
axial (A), vector (V), scalar (S), and tensor (T) charges of the
nucleon~\cite{Gupta:2018qil, Bhattacharya:2016zcn}.
On the lattice, the nucleon charges are extracted from the ratio
of the three-point [$C_\textrm{3pt}^{A,S,T,V}(\tau,t)$] to 
two-point [$C_\textrm{2pt}(\tau)$] correlation functions of the nucleon. 
In the three-point function, a quark bilinear operator $\bar{q} \Gamma q$ is 
inserted at Euclidean time $t$ between the nucleon source and sink. The desired ground-state result is 
obtained by removing the excited-state
contamination~\cite{Bhattacharya:2011qm,Bhattacharya:2015esa} using 
calculations at multiple source-sink separations, $\tau$, 
and extrapolating the results to $\tau\rightarrow\infty$.

The results presented use correlations functions already calculated on
the $a09m310$ ensemble generated by the MILC
Collaboration~\cite{Follana:2006rc, Bazavov:2012xda} at lattice
spacing $a \approx 0.089$~fm and pion mass $M_\pi \approx
313$~MeV~\cite{Gupta:2018qil,Bhattacharya:2016zcn}.  The data consist
of 144,832 measurements on 2263 gauge configurations. On each
configuration, 64 measurements from randomly chosen and widely
separated source positions were made. The quark propagators were
calculated using the Multigrid
inverter~\cite{Babich:2010qb,Osborn:2010mb}, ported in the CHROMA
software suite~\cite{Edwards:2004sx}, with a sloppy stopping criterion. 
The bias introduced by using a sloppy convergence condition is much 
smaller than the statistical uncertainty for nucleon
observables~\cite{Bhattacharya:2016zcn,Yoon:2016dij} and, is therefore,
neglected in this study. If necessary, however, it can be easily 
incorporated by modifying Eq.~\eqref{eq:o_unbiased}.

The correlation coefficients between the various $C_\textrm{3pt}$
measured at \(t=\tau/2=5a\) and the $C_\textrm{2pt}$ at various values
of $\tau$, are shown in Fig.~\ref{fig:corr_map}.  The strongest
correlation is with the value of $C_\textrm{2pt}$ near the sink of
$C_\textrm{3pt}$ at $\tau=10a$, and not near the $t=5a$ of operator
insertion.  Our intuitive understanding of why the correlation is
strongest with $C_\textrm{2pt}(10a)$ is as follows: the spectral
decompositions of the two correlation functions are similar except for
the insertion of the operator at $t=5a$ in $C_\textrm{3pt}(10a)$. If
the ground state saturates these correlation functions, then the extra
term in $C_\textrm{3pt}$ is the matrix element of this operator within
the ground state of the proton.  This matrix element can be considered
as inserting a number (the charge) at $t=5a$ in $C_\textrm{2pt}$. 
If the configuration to configuration fluctuations in the matrix element 
are small, then one expects a strong correlation between
$C_\textrm{2pt}(10a)$ and $C_\textrm{3pt}(10a)$.  In addition, there
are strong correlations between successive time slices of
$C_\textrm{2pt}$; thus, one expects the correlation of
$C_\textrm{3pt}(10a)$ with $C_\textrm{2pt}$ to be spread over a few
time slices about $t=10a$ as also indicated by the data in
Fig.~\ref{fig:corr_map}.  In the more realistic case, in which the
nucleon wave function at $t=5a$ has significant contributions from a
tower of excited states, the operator can also cause transitions
between these states, and its insertion can no longer be approximated
by just one number.  One can still expect that operators for which these
transition matrix elements are small will have stronger correlations.
Based on the observed pattern of excited states, discussed in
Ref.~\cite{Gupta:2018qil}, we expect the ordering of
correlations $V > T > A > S$, whereas the observed pattern shown in
Fig.~\ref{fig:corr_map} is $V > S > T > A $.  \looseness-1

It is the existence of such correlations that allows the prediction of
$C_\textrm{3pt}$ from $C_\textrm{2pt}$ using a boosted decision tree
(BDT) regression algorithm available in SCIKIT-LEARN PYTHON ML
library~\cite{scikit-learn}.  BDT is a ML algorithm that builds an
ensemble (tower) of simple decision trees such that each successive
decision tree corrects the prediction error of the previous decision
tree. The result is a powerful regression algorithm with small number
of tuning parameters and a low risk of overfitting.  It is also fast:
for the data sizes we are considering, it only takes a couple of
minutes on a laptop to find an appropriate predictor and evaluate it
on the unlabeled samples.  The SCIKIT-LEARN implementation of the BDT
we used in this study is based on the Classification and Regression
Trees (CART) algorithm~\cite{breiman1984classification} with gradient
boosting~\cite{Friedman00greedyfunction,Friedman:2002:SGB:635939.635941}.
For the prediction of $C_\textrm{3pt}$, we use 100 boosting stages of
depth-3 trees with learning rate of 0.1 and a subsampling of 0.7.
Note that, in this example, the pattern of correlation is such that a
linear regression algorithm (such as LASSO
\cite{Santosa:1986,Tibshirani94regressionshrinkage} or
RIDGE~\cite{Hoerl1}) gives predictions with reasonable precision. Such
a simplification does not occur for the second example described
later.

\begin{table}[tbp]
\begin{ruledtabular}
\begin{tabular}{lllll}
$\Gamma$ & DM  & ${\cal P}1$  & \quad ~Bias    & Prediction  \\ 
         &     &              &                & without BC  \\ \hline
S & 0.936(10)  & 0.933(15)    & $+$0.002(46)   & 0.934(14)   \\
A & 1.2011(41) & 1.1997(48)   & $-$0.0003(105) & 1.1999(46)  \\
T & 1.0627(34) & 1.0638(39)   & $-$0.0004(78)  & 1.0636(38)  \\
V & 1.0462(36) & 1.0455(36)   & $+$0.0002(20)  & 1.0456(36)  \\
\end{tabular}
\caption{Average of $C_\textrm{3pt}^{\Gamma}(10a, 5a) / \langle
  C_\textrm{2pt}(10a)\rangle$ on the unlabeled dataset. DM is
  the directly measured result, ${\cal P}1$ is the BC prediction
  defined in the text, with the bias correction factor given in column
  4. For the prediction without BC, we used the full 680 labeled
  configurations for training of the BDT.  Note that for this large
  dataset, the bias correction and the increase in the error in the
  prediction with BC are negligible.
\label{tab:pred_3pt}
}
\end{ruledtabular}
\end{table}

The outline of the calculation is as follows:
\begin{enumerate}
\item
For each $(\tau,t)$, the BDT is trained using the set of
$C_\textrm{2pt}$ data (input) and $C_\textrm{3pt}^{A,S,T,V}(\tau,t)$
(output).  This trained BDT can now take as input the unseen
$C_\textrm{2pt}$ data and output the predicted
$C_\textrm{3pt}^{A,S,T,V}(\tau,t)$. To predict $C_\textrm{3pt}$ at a
given $(\tau,t)$, one can use the data for $C_\textrm{2pt}$ on all
time slices.  The essence of a trained BDT is that it gives larger
weight to the input $C_\textrm{2pt}$ element with higher correlation with the
target observable.
\item
The trained BDT is first used on the dataset designated for BC data
to predict $C_\textrm{3pt}^{A,S,T,V}(\tau,t)$.  The bias correction
factor is then determined by comparing this prediction with the
corresponding directly measured value on the same BC set.
\item
The trained BDT is next used on the unlabeled $C_\textrm{2pt}$ dataset to give the predicted 
$C_\textrm{3pt}^{A,S,T,V}(\tau,t)$. 
\item
To the average of this predicted $C_\textrm{3pt}^{A,S,T,V}(\tau,t)$ set, the bias correction factor 
is added to give the BC prediction we call ${\cal{P}} 1$. 
\item
The statistical precision can be improved by constructing the weighted
average of the BC prediction ${\cal{P}} 1$ and the direct measured (DM) results on the labeled dataset. We
call this estimate ${\cal{P}} 2$. Note that the direct measurements on
the labeled data and the predictions on the unlabeled data are not
identically distributed because the prediction is not exact, however,
the bias-corrected mean is the same. Therefore, when performing
excited-state fits discussed below, we simultaneously fit the two data
sets with common fit parameters.
\end{enumerate}
The training and prediction steps treat data from each source position
as independent, whereas the bias-corrected estimates for each
bootstrap sample are obtained using configuration averages in
Eq.~\eqref{eq:o_unbiased}. In this case, the errors are obtained using
500 bootstrap samples.

For the first example, we choose 680 of the 2263 configurations,
separated by three configurations in trajectory order, as the labeled
data.  To determine the number of configurations to use for training,
we varied the number between 30 and 180. We found that the variance of
the prediction on the unlabeled dataset was the smallest and roughly
constant between 60 and 120. We, therefore, picked 60 configurations
from the labeled set for training and 620 for bias correction.  The
1583 unlabeled configurations were used for prediction.  The BDT
regression algorithm was trained to predict
$C_\textrm{3pt}^{A,S,T,V}(\tau,t)/{\cal N}$ for all $\tau$ and $t$
with $\{C_\textrm{2pt}(\tau)/{\cal N} \textrm{ for }
\tau/a=0,1,2,\ldots,20\}$ as input. The normalization ${\cal N} \equiv
        {\langle{C_\textrm{2pt}(\tau)}\rangle}$ was needed to make
        numbers of $O(1)$ for numerical stability of the BDT in the
        SCIKIT-LEARN library.

\begin{figure}[tbp]
\includegraphics[width=0.48\textwidth]{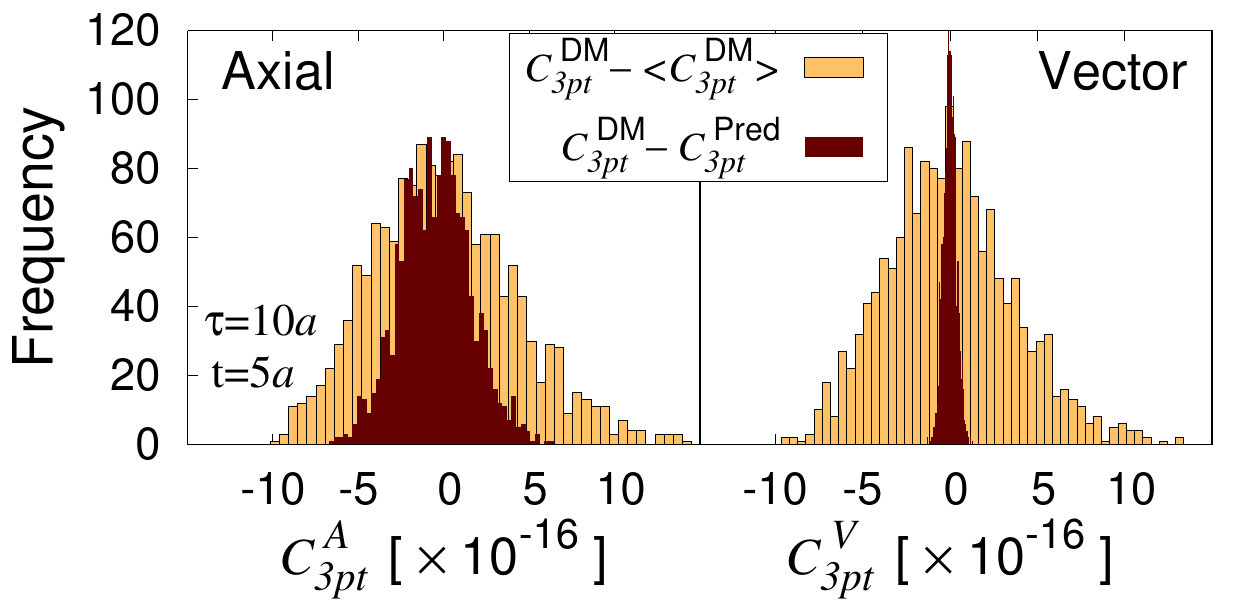}\\
\includegraphics[width=0.48\textwidth]{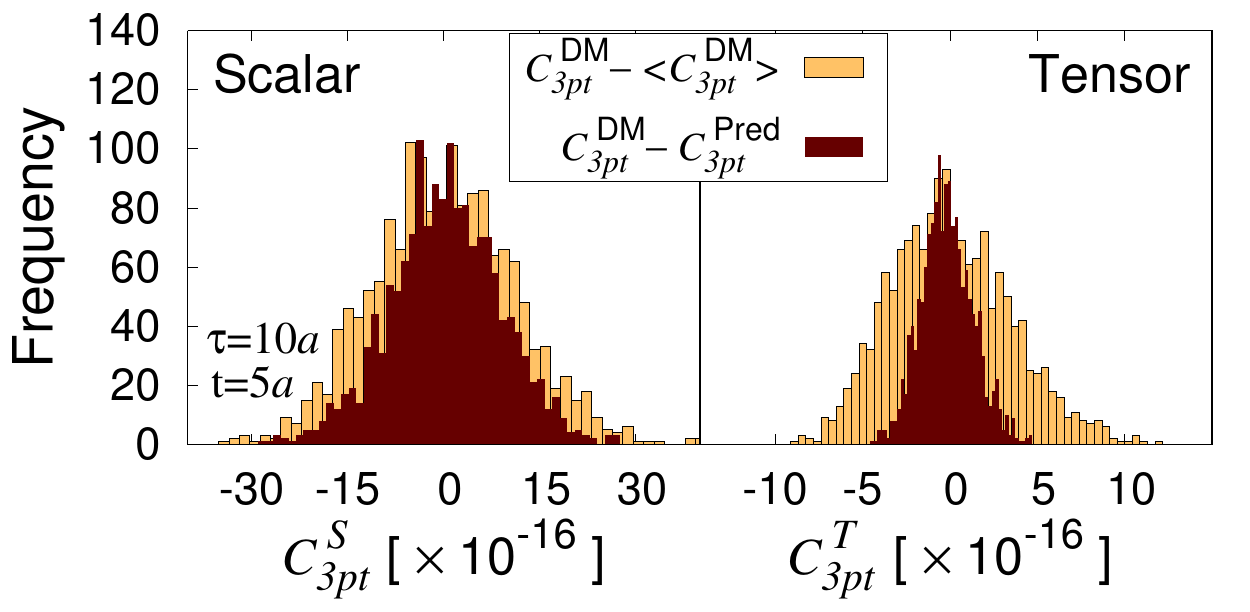}
\caption{Statistical distribution of $C_\textrm{3pt}(10a, 5a)$ (light gold) and the prediction error (dark red).
The ratios of the standard deviations of the prediction error (PE) and DM data
at $t=\tau/2=5a$ are
${\sigma_{\textrm{PE}}/\sigma_{\textrm{DM}}}$~=~0.79, 0.49, 0.44 and
0.12 for S, A, T and V, respectively. 
\label{fig:dist_gA_gV}}
\end{figure}

Data in Table~\ref{tab:pred_3pt} show that the statistical errors in
the bias correction term are large; however, the error
in the BC estimate is essentially identical to that
in the DM estimates. This implies strong
correlations between the two terms, uncorrected and the BC factor.
Fig.~\ref{fig:dist_gA_gV} shows that the statistical fluctuations in
the DM data are larger than the prediction error (PE~$\equiv
C_\textrm{3pt}^{\rm DM} - C_\textrm{3pt}^{\rm Pred}$) of the ML
algorithm. The ratios of the standard deviations,
${\sigma_{\textrm{PE}}/\sigma_{\textrm{DM}}}$, of the PE and DM data
are given in Table~\ref{tab:pred_qual}. This pattern of smaller
variance leads us to believe that, with further optimization, the
reduction in computation cost given in Table~\ref{tab:extrap} can be
increased significantly.  \looseness-1

\begin{table}[tbp]
\begin{ruledtabular}
\begin{tabular}{lllll}
\multicolumn{5}{l}{${\sigma_{\textrm{PE}}/\sigma_{\textrm{DM}}}$ of ${\cal P}2$} \\ \hline
$\Gamma$ & $\tau=8$  & $\tau=10$ & $\tau=12$ & $\tau=14$ \\ \hline
S        & 0.791(16) & 0.793(15) & 0.791(14) & 0.785(14) \\
A        & 0.394(9)  & 0.493(12) & 0.601(13) & 0.721(14) \\
T        & 0.334(9)  & 0.439(11) & 0.571(13) & 0.705(14) \\
V        & 0.089(4)  & 0.115(8)  & 0.134(7)  & 0.159(6)  \\\hline\hline
\multicolumn{5}{l}{${\sigma_{\textrm{PE}}/\sigma_{\textrm{DM}}}$ of ${\cal VP}2$} \\ \hline
$\Gamma$ & $\tau=8$  & $\tau=10$ & $\tau=12$ & $\tau=14$ \\ \hline
S        & 0.696(14) & 0.535(12) & input     & 0.546(12) \\
A        & 0.357(9)  & 0.355(9)  & input     & 0.501(12) \\
T        & 0.304(8)  & 0.329(9)  & input     & 0.498(12) \\
V        & 0.089(5)  & 0.105(10) & input     & 0.143(7)  \\
\end{tabular}
\caption{Ratio ${\sigma_{\textrm{PE}}/\sigma_{\textrm{DM}}}$
  representing quality of the prediction ${\cal P}2$ (upper) and
  ${\cal VP}2$ (lower) at $t=\tau/2$. Smaller values indicate
  better prediction.
\label{tab:pred_qual}
}
\end{ruledtabular}
\end{table}

We have carried out two kinds of tests of the efficacy of the
method. In Table~\ref{tab:pred_3pt_2}, we show data for
$C_\textrm{3pt}^{\Gamma}(10a, 5a) / \langle
C_\textrm{2pt}(10a)\rangle$ for different numbers of labeled data, 
keeping (i) the full $2263$ configurations and (ii) 500
configurations. We find that the results are consistent for different
numbers, \emph{Prediction($N,N_t$)}, of labeled data in both cases.
Even when only ten configurations (640 measurements) are used for the
training dataset, one gets reasonable estimates.  The errors scale
roughly as the total number of configurations as can be seen by
comparing the upper and lower tables.

\begin{table}[tbp]
\begin{ruledtabular}
\begin{tabular}{lllll}
\multicolumn{5}{l}{Total number of configurations: 2263} \\ \hline
$\Gamma$ & DM  & Prediction & Prediction & Prediction  \\ 
         &     & (680,60)   &  (450,40)  & (225,20)    \\ \hline
S & 0.930(09)  & 0.925(14)  & 0.937(19)  & 0.959(26)  \\
A & 1.1984(33) & 1.1967(42) & 1.1991(52) & 1.2046(69) \\
T & 1.0611(27) & 1.0615(35) & 1.0637(40) & 1.0659(51) \\
V & 1.0437(28) & 1.0427(28) & 1.0437(31) & 1.0434(32) \\ \hline\hline
\multicolumn{5}{l}{Total number of configurations: 500} \\ \hline
$\Gamma$ & DM  & Prediction & Prediction & Prediction  \\ 
         &     & (150,20)   &  (100,20)  & (50,10)    \\ \hline
S &  0.935(19) &  0.904(29) &  0.909(32) &  0.988(40) \\
A & 1.1940(77) & 1.1848(99) &  1.188(11) &  1.191(17) \\
T & 1.0588(62) & 1.0663(78) & 1.0555(88) & 1.0495(112) \\
V & 1.0437(64) & 1.0429(64) & 1.0417(63) & 1.0443(70) \\
\end{tabular}
\caption{Average of $C_\textrm{3pt}^{\Gamma}(10a, 5a) / \langle
  C_\textrm{2pt}(10a)\rangle$ on the full dataset of 2263 (upper) and
  500 (lower) configurations. \emph{Prediction($N,N_t$)} denotes 
  predictions made using $N$ labeled configurations of which $N_t$ are
  used for training.
\label{tab:pred_3pt_2}
}
\end{ruledtabular}
\end{table}

\begin{figure*}[tbp]
\includegraphics[width=0.98\textwidth]{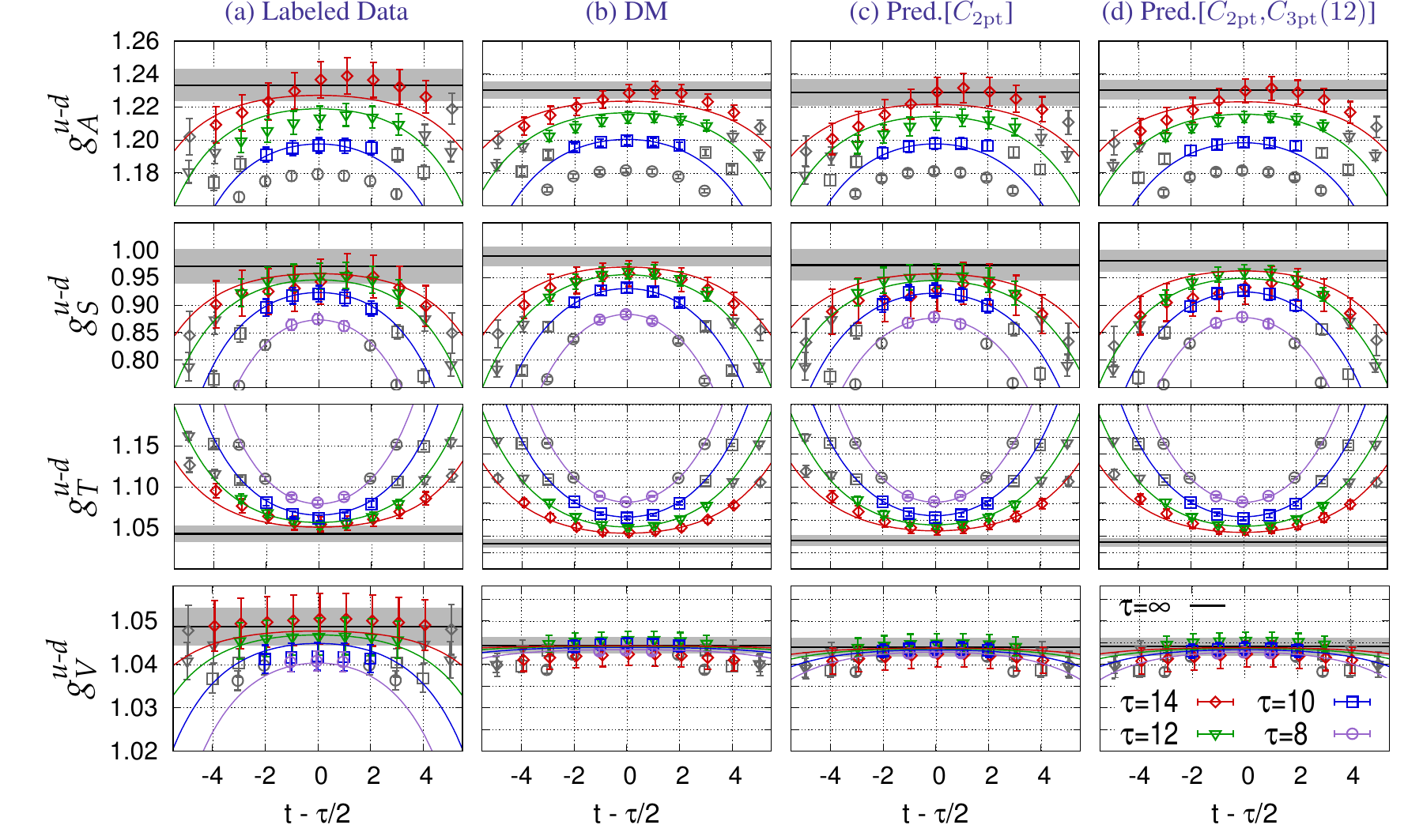}
\caption{Removing excited-state contamination using the two-state fit
  for (a) DM on the labeled data, (b)
  DM on full data, (c) DM on labeled data combined with ML
  predictions from $C_\textrm{2pt}$ on unlabeled data (${\cal{P}} 2$), and (d) DM on
  labeled data combined with ML predictions from $C_\textrm{2pt}$ and
  $C_\textrm{3pt}(\tau=12a)$ on unlabeled data (${\cal{VP}} 2$).
\label{fig:pred_g}}
\end{figure*}

In Fig.~\ref{fig:pred_g}, we compare the prediction ${\cal{P}} 2$ of
$C_\textrm{3pt}^{A,S,T,V}$ at all $\tau$ and $t$ [column (c)] with
the DM on labeled and full data shown in columns (a) and (b),
respectively.  The observed dependence on $\tau$ and $t$ is due to
contributions from excited states of the nucleon, and the desired
ground-state result is given by the limit $\tau \to \infty$. This can
be obtained by fitting the data at various $t$ and $\tau$ using the
spectral decomposition of
$C_\textrm{3pt}^{A,S,T,V}$. Fig.~\ref{fig:pred_g} shows such a fit
assuming only the lowest two states contribute to the spectral
decomposition, i.e., the two-state fit described
in Refs.~\cite{Bhattacharya:2013ehc,Gupta:2018qil,Bhattacharya:2016zcn}.
The lines show the result of this fit for the various $\tau$, and the
gray band gives the $\tau \to \infty$ value. We find that the
prediction ${\cal{P}} 2$ in column (c) is consistent with the DM
results on the full dataset. 

We can further improve the prediction if data for a single value of
$\tau$, say $C_\textrm{3pt}^{A,S,T,V}(\tau/a=12)$, is available on the
full dataset. Then, in the training stage, we use as input both
$C_\textrm{2pt}$ and $C_\textrm{3pt}^{A,S,T,V}(\tau/a=12)$.
Having trained the BDT on the labeled data, we now use
$C_\textrm{2pt}$ and $C_\textrm{3pt}^{A,S,T,V}(\tau/a=12)$ as
input to predict $C_\textrm{3pt}(\tau/a=8,10,14)$, which we label
${\cal{VP}} 2$.  These results are shown in Fig.~\ref{fig:pred_g}
column (d). Including $C_\textrm{3pt}^{A,S,T,V}(\tau/a=12)$ in the training and the prediction stages 
increases the computational cost relative to ${\cal{P}} 2$ but reduces the errors. 
For a fixed size of error, ${\cal{VP}} 2$ is more efficient than ${\cal{P}} 2$ as shown in 
Table~\ref{tab:extrap}. 

A comparison of the predictions from
$C_\textrm{2pt}$  (${\cal P}2$)  and from $C_\textrm{2pt}$ and
$C_\textrm{3pt}^{A,S,T,V}(\tau/a=12)$  (${\cal VP}2$) vs DM is 
shown in Table~\ref{tab:extrap} for the charges $g_{A,S,T,V}$ obtained after the 
extrapolation $\tau \to \infty$ using the four values of $\tau$. While both estimates, 
${\cal P}2$ and ${\cal VP}2$, are consistent with the DM, ${\cal VP}2$ is closer to 
DM with respect to both the central value and the error. 
Taking into account the increase in the statistical uncertainty
(scaling the cost by the square of the number of measurements) in the
predicted results, the ML analysis ${\cal VP}2$ provides between
a 7\% and 26\% reduction in the computational cost.  The amount of gain is
observable dependent.

\begin{table}[tbp]
\begin{ruledtabular}
\begin{tabular}{l|l|lr|lr}
      & DM    & ${\cal{P}}2 (\tau\to\infty)$ & Cost(\%) & ${\cal{VP}}2 (\tau\to\infty)$  & Cost(\%) \\
      &       & \multicolumn{2}{l|}{[$C_\textrm{2pt}$]} & \multicolumn{2}{l}{[$C_\textrm{2pt}$,$C_\textrm{3pt}$(12)]}\\
\hline
$g_S$ & 0.989(18)  & 0.973(29)  & 138 & 0.981(20) & 80   \\
$g_A$ & 1.2303(51) & 1.2289(83) & 141 & 1.2304(61)& 93   \\
$g_T$ & 1.0311(51) & 1.0347(68) & 97  & 1.0326(54)& 74   \\
$g_V$ & 1.0443(19) & 1.0439(22) & 74  & 1.0440(21)& 78   \\
\end{tabular}
\end{ruledtabular}
\caption{Comparison of \(\tau\to\infty\) extrapolated nucleon charges
  calculated from the ML predictions ${\cal{P}}2$ and ${\cal{VP}}2$
  and the relative computational cost vs the DM.  The cost
  includes the factor required to make the errors the same assuming
  they scale as $M^2$. }
\label{tab:extrap}
\end{table}


\section{Experiment B: CP violating phase in the Neutron State}
\label{sec:EXPb}
The second example is taken from the calculation of the matrix element
of the chromoelectric dipole moment (cEDM) operator,
${O_{\textrm{cEDM}}{\equiv}i\bar{q}(\sigma_{\mu\nu}G^{\mu\nu})\gamma_5
  q}$ where $G^{\mu\nu}$ is the gluon field strength tensor, within
the neutron state.  It arises in theories beyond the standard model
and violates parity P and time-reversal T symmetries, or
equivalently, charge C and CP symmetries in theories invariant under
CPT. Since any CP violating (CPV) operator gives a contribution to the
neutron electric dipole moment (nEDM), a bound or a nonzero value for nEDM in coming
experiments will constrain novel CP violation \cite{Pospelov:2005pr, RamseyMusolf:2006vr, Engel:2013lsa}.
So far only preliminary
lattice QCD calculations exist and cost-effectively improving the statistical signal is
essential~\cite{Syritsyn:2018mon,Kim:2018rce,Bhattacharya:2018qat}. 
We have proposed a Schwinger source method approach
(SSM)~\cite{Bhattacharya:2016oqm,Bhattacharya:2016rrc} that exploits
the fact that the cEDM operator is a quark bilinear. In the SSM, 
effects of the cEDM interaction are incorporated into the two- and 
three-point functions by modifying the Dirac clover fermion action: 
\begin{align}
D_{\text{clov}} & \rightarrow D_{\text{clov}} + i \varepsilon \sigma_{\mu\nu} \gamma_5   G^{\mu\nu} \nonumber \\
D_{\text{clov}} & \rightarrow D_{\text{clov}} + i \varepsilon_5   \gamma_5 \,.
\end{align}
The second equation is for the pseudoscalar operator $O_{\gamma_5}
{\equiv}i\bar{q} \gamma_5 q$ that mixes with cEDM due to quantum
effects~\cite{Bhattacharya:2015rsa}.  

With CP violation, the Dirac equation for the neutron spinor $u$
becomes $(ip_\mu\gamma_\mu+me^{-i 2 \alpha\gamma_5})u=0$; i.e., the
neutron spinor acquires a CP-odd phase $\alpha$ ($\alpha_5$), which is
expected to be linear in $\varepsilon$ ($\varepsilon_5$) for small
$\varepsilon$ ($\varepsilon_5$). At leading order, these phases can be
obtained from the four two-point functions, $C_{\text{2pt}}$,
$C_{\text{2pt}}^P$, $C_{\text{2pt}}^{P,\varepsilon}$, and
$C_{\text{2pt}}^{P,\varepsilon_5}$, where the superscript $P$
indicates an additional factor of $\gamma_5$ is included in the spin
projection~\cite{Shintani:2015vsx,Bhattacharya:2018qat}.\footnote{$C_{\text{2pt}}^P$
  has a zero mean but fluctuations correlated with
  $C_{\text{2pt}}^{P,\varepsilon}$ and
  $C_{\text{2pt}}^{P,\varepsilon_5}$. It can, therefore, be used for
  variance reduction~\protect\cite{Bhattacharya:2018qat}.}  The
correlator $C_{\text{2pt}}^{P,\varepsilon}$ (
$C_{\text{2pt}}^{P,\varepsilon_5}$) is constructed using quark
propagators with the $O_{\textrm{cEDM}}$ ($O_{\gamma_5}$) term and is
expected to be imaginary and vanish as $\varepsilon \to 0$
($\varepsilon_5 \to 0$). In a first step, we show predictions of the
BDT regression algorithm for these two using only $C_{\text{2pt}}$ and
$C_{\text{2pt}}^P$.

\begin{figure}[tbp]
\includegraphics[width=0.48\textwidth]{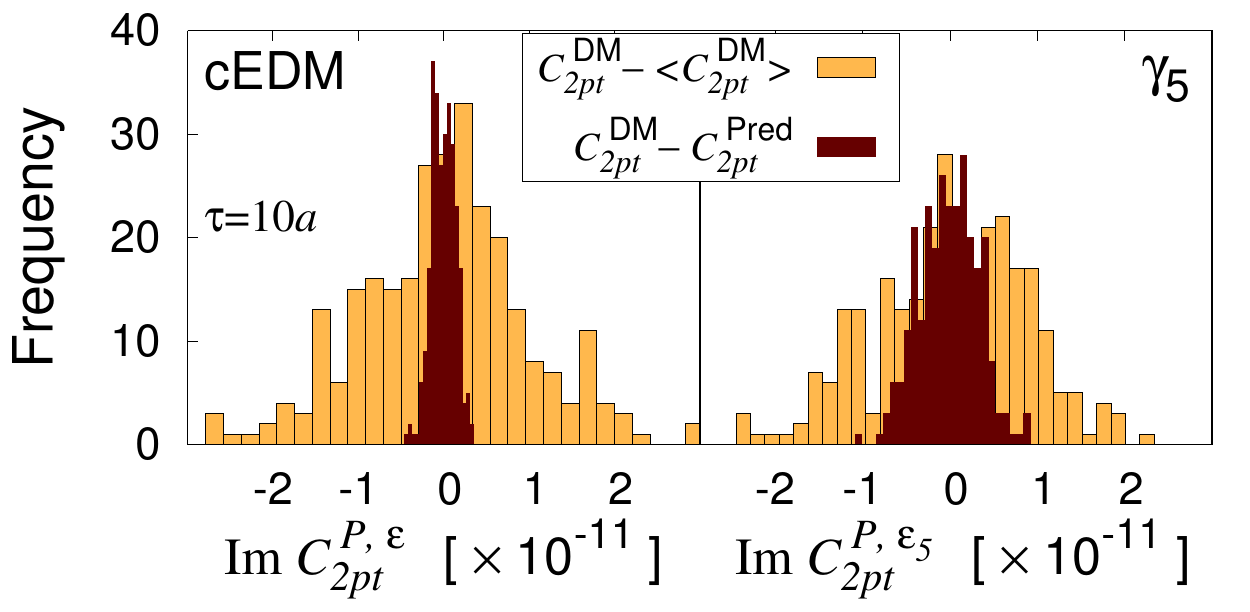}
\caption{Distribution of
  $\operatorname{Im}\big[C_\textrm{2pt}^{P,\varepsilon}(10a)\big]$
  (left) and
  $\operatorname{Im}\big[C_\textrm{2pt}^{P,\varepsilon_5}(10a)\big]$
  (right), averaged over sources in each configuration, are shown in
  light gold and the prediction error in dark red. The ratio of the
  standard deviations $\sigma_{\textrm{PE}}/\sigma_{\textrm{2pt}}
  \approx$ 0.18 for $O_{\textrm{cEDM}}$ and 0.4 for $O_{\gamma_5}$.
\label{fig:dist_cedm}}
\end{figure}

For the training and prediction, we use the $C_{\text{2pt}}$,
$C_{\text{2pt}}^P$, $C_{\text{2pt}}^{P,\varepsilon}$ and
$C_{\text{2pt}}^{P,\varepsilon_5}$ measured in
Refs.~\cite{Bhattacharya:2016oqm,Bhattacharya:2016rrc} on 400 MILC
highly improved staggered quarks (HISQ) lattices at $a=0.12$~fm and $M_\pi=310$~MeV (the $a12m310$
ensemble) with clover fermions. On each configuration, these
correlators are constructed using 64 randomly chosen widely separated
sources with a sloppy stopping condition, the effects of which are again
ignored. Out of the 400 configurations, 120 configurations, separated
by three configurations in trajectory order, are chosen as the labeled
data, and the remaining 280 configurations are used as the unlabeled
data. From the labeled data, 70 randomly chosen configurations are
used for training.  Only 50 configurations sufficed for bias
correction in this case because the ratio of standard deviations of
the prediction error vs the DM
(${\sigma_{\textrm{PE}}/\sigma_{\textrm{DM}}}$) is small as shown in
Fig.~\ref{fig:dist_cedm}. Errors are obtained using 200 bootstrap samples.

The BDT regression algorithm is trained to predict the imaginary parts
of $C_{\text{2pt}}^{P,\varepsilon}$ and
$C_{\text{2pt}}^{P,\varepsilon_5}$ using both the real and imaginary
parts of $C_{\text{2pt}}$ and $C_{\text{2pt}}^P$. Note that in the
absence of the CPV terms, $C_{\text{2pt}}^P$ and the imaginary part of
$C_{\text{2pt}}$ average to zero, but, they have nonzero correlations
with the target imaginary parts of $C_{\text{2pt}}^{P,\varepsilon}$
and $C_{\text{2pt}}^{P,\varepsilon_5}$. The BDT regression algorithm
with 500 boosting stages of depth-3 trees with learning rate of 0.1
and subsampling of 0.7 gives a good prediction as shown in
Fig.~\ref{fig:dist_cedm}. Because of nonlinear correlations, the BDT
works better than linear regression algorithms in this case; the
prediction error is about 50\% larger with linear models at $t=1$ and
decreases to less than 10\% by $t=8$.  Again, for numerical stability,
all data fed into the BDT algorithm are normalized by
$\langle{C_{\text{2pt}}(\tau)}\rangle$.

\begin{figure}[tbp]
\includegraphics[width=0.48\textwidth]{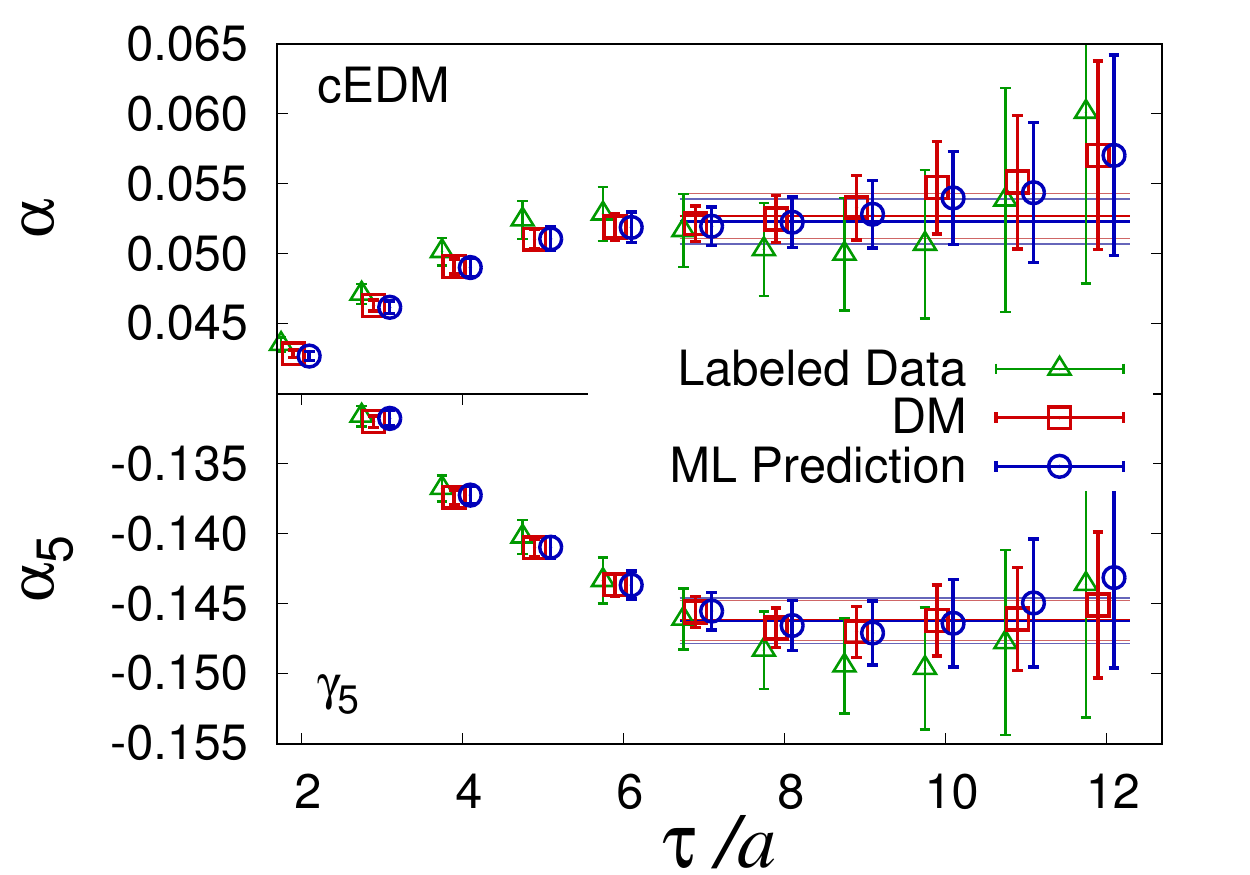}
\caption{CPV phase $\alpha$ calculated from the DM
  $C_{\text{2pt}}^P$ on the full data (red squares), improved ML prediction (blue circles), 
  and the labeled data (green triangles).
\label{fig:alpha}}
\end{figure}

Using the predicted $C_{\text{2pt}}^{P,\varepsilon}$ and
$C_{\text{2pt}}^{P,\varepsilon_5}$ on all time slices, we
calculate the CPV phases $\alpha$ and $\alpha_5$ by taking their ratio
with $C_{\text{2pt}}$, because $C_{\text{2pt}}^{\varepsilon,\varepsilon_5}$ differ
from $ C_{\text{2pt}}$ at $O(\varepsilon^2)$. Fig.~\ref{fig:alpha} shows the comparison between
the CPV phase calculated from DM on the full and labeled data and the ML
predicted data. The horizontal lines give the averages over the plateau region where 
the excited-state contamination is small. Results for $\alpha$ and $\alpha_5$ 
 are summarized in Table~\ref{tab:alpha_plateau}. To get the improved ML predictions, we 
combine the prediction on the 280 unlabeled configurations with the 
DM data on the 120 labeled configurations. These combined data are 
analyzed following the same bootstrap resampling procedure used in the first example discussed earlier. 

The prediction uses 30\% of the data for
$C_{\text{2pt}}^{P,\varepsilon}$ and
$C_{\text{2pt}}^{P,\varepsilon_5}$ and 100\% for $C_{\text{2pt}}^P$
  and $ C_{\text{2pt}}$. This reduces the total number of propagator
calculations by 47\% compared to the direct measurement. Taking into
account the increase of the statistical uncertainty, the
computational cost reduction is about 30\% as shown in Table~\ref{tab:alpha_plateau}.

\begin{table}[tbp]
\begin{ruledtabular}
\begin{tabular}{l|c|cr}
      & DM    & ${\cal{P}}2 $ & Cost  \\
\hline
$\alpha$ & \phantom{$+$}0.0527(17) & \phantom{$+$}0.0525(18)  & 62\%\\
$\alpha_{5}$      & $-$0.1463(14)  & $-$0.1460(17)            & 77\%\\
\end{tabular}
\end{ruledtabular}
\caption{Comparison of the ML prediction of the CPV phases $\alpha$ and $\alpha_5$ and the relative cost 
vs the DM results.}
\label{tab:alpha_plateau}
\end{table}

\section{Conclusion}

In conclusion, the proposed ML algorithm used to predict
compute-intensive observables from simpler measurements gives a modest
computational cost reduction of 7\%--38\% depending on the observables
analyzed here as summarized in Tables~\ref{tab:extrap} (${\cal{VP}}2$)
and~\ref{tab:alpha_plateau} (${\cal{P}}2$).  The technique is, however, general
provided one can find inexpensive measurements that correlate well
with the observable of interest.  The computational cost reduction
depends on the degree of correlations.  We are investigating other ML
methods to further improve the quality of the prediction and reduce
computational cost.\looseness-1


\begin{acknowledgments}
We thank the MILC Collaboration for providing the 2+1+1-flavor 
highly improved staggered quarks lattices. Simulations were carried 
out on computer facilities at (i) the National Energy Research 
Scientific Computing Center, a DOE Office of Science User Facility 
supported by the Office of Science of the U.S. Department of Energy 
under Contract No. DE-AC02-05CH11231; (ii) the Oak Ridge Leadership 
Computing Facility at the Oak Ridge National Laboratory, which is 
supported by the Office of Science of the U.S. Department of Energy 
under Contract No. DE-AC05-00OR22725; (iii) the USQCD Collaboration, 
which is funded by the Office of Science of the U.S. Department of 
Energy; and (iv) Institutional Computing at Los Alamos National 
Laboratory. Authors were supported by the U.S. Department of Energy, 
Office of Science, Office of High Energy Physics under Contract No. 
89233218CNA000001 and by the Los Alamos National Laboratory (LANL) 
LDRD program. B.~Y. acknowledges support from the U.S. Department 
of Energy, Office of Science, Office of Advanced Scientific 
Computing Research and Office of Nuclear Physics, Scientific Discovery 
through Advanced Computing (SciDAC) program.
\end{acknowledgments}

\appendix
\section{Example python code for machine learning regression}
A simplified example of the python code used for ML training and
prediction, using a BDT regression algorithm provided by the
SCIKIT-LEARN library~\cite{scikit-learn}, is given in
Listing~\ref{list:code}. The code shows the calls for importing the
SCIKIT-LEARN module, defining and training the BDT regressor, making
predictions using the trained regressor, and implementing the 
BC procedure.  Here we assumed that the required
training, BC, and unlabeled data are given by a LOAD\_DATA()
function. The bootstrap procedure, needed to estimate uncertainty of
the final BC prediction, is implicitly wrapped around the various
calls in the code in Listing~\ref{list:code} as described in
Section~\ref{sec:intro}.

\definecolor{codegreen}{rgb}{0,0.6,0}
\definecolor{codegray}{rgb}{0.5,0.5,0.5}
\definecolor{codepurple}{rgb}{0.58,0,0.82}
\definecolor{backcolour}{rgb}{0.95,0.95,0.92}
 
\lstset{
  backgroundcolor=\color{backcolour},
  commentstyle=\color{codegreen},
  keywordstyle=\color{magenta},
  numberstyle=\tiny\color{codegray},
  stringstyle=\color{codepurple},
  basicstyle=\footnotesize,
  breakatwhitespace=false,        
  breaklines=true,                
  captionpos=b,                   
  keepspaces=true,                
  numbers=left,                   
  numbersep=5pt,                 
  showspaces=false,               
  showstringspaces=false,
  showtabs=false,                 
  tabsize=2,
  xleftmargin=0.5cm}
 
\begin{widetext}
\begin{lstlisting}[language=Python,
label={list:code},
caption={Python example code for calculating bias-corrected
predictions using BDT in scikit-learn library. In the code,
we assume that the data is given by LOAD\_DATA() function.}]
import numpy as np
from sklearn.ensemble import GradientBoostingRegressor
 
# Load data
# X_i = [o1_i, o2_i, o3_i, ...]; Array of measured observables
# y_i = O_i                    ; DM of target observable O
# X_train, X_bc, X_unlab: arrays of X_i
# y_train, y_bc         : arrays of y_i
X_train, y_train, X_bc, y_bc, X_unlab = LOAD_DATA()
 
# Gradient boosted decision tree regressor
gbr = GradientBoostingRegressor(learning_rate=0.1, n_estimators=100, max_depth=3)
 
# Training regressor to predict y_i from X_i
gbr.fit(X_train, y_train)
 
# Predictions of y on bias correction and unlabeled datasets
y_bc_pred    = gbr.predict(X_bc)
y_unlab_pred = gbr.predict(X_unlab)
 
# Bias correction term
BiasCrxn = np.average(y_bc - y_bc_pred)
 
# Predictions on unlabeled dataset given by the average
PredAvg = np.average(y_unlab_pred)
 
# Bias-corrected prediction of <O>
BC_Pred = PredAvg + BiasCrxn
\end{lstlisting}
\end{widetext}

\bibliographystyle{apsrev}
\bibliography{refs}

\begin{thebibliography}{38}
\expandafter\ifx\csname natexlab\endcsname\relax\def\natexlab#1{#1}\fi
\expandafter\ifx\csname bibnamefont\endcsname\relax
  \def\bibnamefont#1{#1}\fi
\expandafter\ifx\csname bibfnamefont\endcsname\relax
  \def\bibfnamefont#1{#1}\fi
\expandafter\ifx\csname citenamefont\endcsname\relax
  \def\citenamefont#1{#1}\fi
\expandafter\ifx\csname url\endcsname\relax
  \def\url#1{\texttt{#1}}\fi
\expandafter\ifx\csname urlprefix\endcsname\relax\def\urlprefix{URL }\fi
\providecommand{\bibinfo}[2]{#2}
\providecommand{\eprint}[2][]{\url{#2}}

\bibitem[{\citenamefont{Wilson}(1974)}]{Wilson:1974sk}
\bibinfo{author}{\bibfnamefont{K.~G.} \bibnamefont{Wilson}},
  \bibinfo{journal}{Phys. Rev.} \textbf{\bibinfo{volume}{D10}},
  \bibinfo{pages}{2445} (\bibinfo{year}{1974}), \bibinfo{note}{[,319(1974)]}.

\bibitem[{\citenamefont{Creutz}(1980)}]{Creutz:1980zw}
\bibinfo{author}{\bibfnamefont{M.}~\bibnamefont{Creutz}},
  \bibinfo{journal}{Phys. Rev.} \textbf{\bibinfo{volume}{D21}},
  \bibinfo{pages}{2308} (\bibinfo{year}{1980}).

\bibitem[{\citenamefont{Bali et~al.}(2010)\citenamefont{Bali, Collins, and
  Schafer}}]{Bali:2009hu}
\bibinfo{author}{\bibfnamefont{G.~S.} \bibnamefont{Bali}},
  \bibinfo{author}{\bibfnamefont{S.}~\bibnamefont{Collins}}, \bibnamefont{and}
  \bibinfo{author}{\bibfnamefont{A.}~\bibnamefont{Schafer}},
  \bibinfo{journal}{Comput. Phys. Commun.} \textbf{\bibinfo{volume}{181}},
  \bibinfo{pages}{1570} (\bibinfo{year}{2010}), \eprint{0910.3970}.

\bibitem[{\citenamefont{Blum et~al.}(2013)\citenamefont{Blum, Izubuchi, and
  Shintani}}]{Blum:2012uh}
\bibinfo{author}{\bibfnamefont{T.}~\bibnamefont{Blum}},
  \bibinfo{author}{\bibfnamefont{T.}~\bibnamefont{Izubuchi}}, \bibnamefont{and}
  \bibinfo{author}{\bibfnamefont{E.}~\bibnamefont{Shintani}},
  \bibinfo{journal}{Phys. Rev.} \textbf{\bibinfo{volume}{D88}},
  \bibinfo{pages}{094503} (\bibinfo{year}{2013}), \eprint{1208.4349}.

\bibitem[{\citenamefont{Baldi et~al.}(2014)\citenamefont{Baldi, Sadowski, and
  Whiteson}}]{Baldi:2014kfa}
\bibinfo{author}{\bibfnamefont{P.}~\bibnamefont{Baldi}},
  \bibinfo{author}{\bibfnamefont{P.}~\bibnamefont{Sadowski}}, \bibnamefont{and}
  \bibinfo{author}{\bibfnamefont{D.}~\bibnamefont{Whiteson}},
  \bibinfo{journal}{Nature Commun.} \textbf{\bibinfo{volume}{5}},
  \bibinfo{pages}{4308} (\bibinfo{year}{2014}), \eprint{1402.4735}.

\bibitem[{\citenamefont{Baldi et~al.}(2015)\citenamefont{Baldi, Sadowski, and
  Whiteson}}]{Baldi:2014pta}
\bibinfo{author}{\bibfnamefont{P.}~\bibnamefont{Baldi}},
  \bibinfo{author}{\bibfnamefont{P.}~\bibnamefont{Sadowski}}, \bibnamefont{and}
  \bibinfo{author}{\bibfnamefont{D.}~\bibnamefont{Whiteson}},
  \bibinfo{journal}{Phys. Rev. Lett.} \textbf{\bibinfo{volume}{114}},
  \bibinfo{pages}{111801} (\bibinfo{year}{2015}), \eprint{1410.3469}.

\bibitem[{\citenamefont{Alexandru et~al.}(2017)\citenamefont{Alexandru,
  Bedaque, Lamm, and Lawrence}}]{Alexandru:2017czx}
\bibinfo{author}{\bibfnamefont{A.}~\bibnamefont{Alexandru}},
  \bibinfo{author}{\bibfnamefont{P.~F.} \bibnamefont{Bedaque}},
  \bibinfo{author}{\bibfnamefont{H.}~\bibnamefont{Lamm}}, \bibnamefont{and}
  \bibinfo{author}{\bibfnamefont{S.}~\bibnamefont{Lawrence}},
  \bibinfo{journal}{Phys. Rev.} \textbf{\bibinfo{volume}{D96}},
  \bibinfo{pages}{094505} (\bibinfo{year}{2017}), \eprint{1709.01971}.

\bibitem[{\citenamefont{Wetzel and Scherzer}(2017)}]{Wetzel:2017ooo}
\bibinfo{author}{\bibfnamefont{S.~J.} \bibnamefont{Wetzel}} \bibnamefont{and}
  \bibinfo{author}{\bibfnamefont{M.}~\bibnamefont{Scherzer}},
  \bibinfo{journal}{Phys. Rev.} \textbf{\bibinfo{volume}{B96}},
  \bibinfo{pages}{184410} (\bibinfo{year}{2017}), \eprint{1705.05582}.

\bibitem[{\citenamefont{Shanahan et~al.}(2018)\citenamefont{Shanahan,
  Trewartha, and Detmold}}]{Shanahan:2018vcv}
\bibinfo{author}{\bibfnamefont{P.~E.} \bibnamefont{Shanahan}},
  \bibinfo{author}{\bibfnamefont{D.}~\bibnamefont{Trewartha}},
  \bibnamefont{and} \bibinfo{author}{\bibfnamefont{W.}~\bibnamefont{Detmold}}
  (\bibinfo{year}{2018}), \eprint{1801.05784}.

\bibitem[{\citenamefont{Efron}(1979)}]{10.2307/2958830}
\bibinfo{author}{\bibfnamefont{B.}~\bibnamefont{Efron}}, \bibinfo{journal}{The
  Annals of Statistics} \textbf{\bibinfo{volume}{7}}, \bibinfo{pages}{1}
  (\bibinfo{year}{1979}), ISSN \bibinfo{issn}{00905364},
  \urlprefix\url{http://www.jstor.org/stable/2958830}.

\bibitem[{\citenamefont{Gupta et~al.}(2018)\citenamefont{Gupta, Jang, Yoon,
  Lin, Cirigliano, and Bhattacharya}}]{Gupta:2018qil}
\bibinfo{author}{\bibfnamefont{R.}~\bibnamefont{Gupta}},
  \bibinfo{author}{\bibfnamefont{Y.-C.} \bibnamefont{Jang}},
  \bibinfo{author}{\bibfnamefont{B.}~\bibnamefont{Yoon}},
  \bibinfo{author}{\bibfnamefont{H.-W.} \bibnamefont{Lin}},
  \bibinfo{author}{\bibfnamefont{V.}~\bibnamefont{Cirigliano}},
  \bibnamefont{and}
  \bibinfo{author}{\bibfnamefont{T.}~\bibnamefont{Bhattacharya}}
  (\bibinfo{year}{2018}), \eprint{1806.09006}.

\bibitem[{\citenamefont{Bhattacharya
  et~al.}(2016{\natexlab{a}})\citenamefont{Bhattacharya, Cirigliano, Cohen,
  Gupta, Lin, and Yoon}}]{Bhattacharya:2016zcn}
\bibinfo{author}{\bibfnamefont{T.}~\bibnamefont{Bhattacharya}},
  \bibinfo{author}{\bibfnamefont{V.}~\bibnamefont{Cirigliano}},
  \bibinfo{author}{\bibfnamefont{S.}~\bibnamefont{Cohen}},
  \bibinfo{author}{\bibfnamefont{R.}~\bibnamefont{Gupta}},
  \bibinfo{author}{\bibfnamefont{H.-W.} \bibnamefont{Lin}}, \bibnamefont{and}
  \bibinfo{author}{\bibfnamefont{B.}~\bibnamefont{Yoon}},
  \bibinfo{journal}{Phys. Rev.} \textbf{\bibinfo{volume}{D94}},
  \bibinfo{pages}{054508} (\bibinfo{year}{2016}{\natexlab{a}}),
  \eprint{1606.07049}.

\bibitem[{\citenamefont{Bhattacharya et~al.}(2012)\citenamefont{Bhattacharya,
  Cirigliano, Cohen, Filipuzzi, Gonzalez-Alonso, Graesser, Gupta, and
  Lin}}]{Bhattacharya:2011qm}
\bibinfo{author}{\bibfnamefont{T.}~\bibnamefont{Bhattacharya}},
  \bibinfo{author}{\bibfnamefont{V.}~\bibnamefont{Cirigliano}},
  \bibinfo{author}{\bibfnamefont{S.~D.} \bibnamefont{Cohen}},
  \bibinfo{author}{\bibfnamefont{A.}~\bibnamefont{Filipuzzi}},
  \bibinfo{author}{\bibfnamefont{M.}~\bibnamefont{Gonzalez-Alonso}},
  \bibinfo{author}{\bibfnamefont{M.~L.} \bibnamefont{Graesser}},
  \bibinfo{author}{\bibfnamefont{R.}~\bibnamefont{Gupta}}, \bibnamefont{and}
  \bibinfo{author}{\bibfnamefont{H.-W.} \bibnamefont{Lin}},
  \bibinfo{journal}{Phys. Rev.} \textbf{\bibinfo{volume}{D85}},
  \bibinfo{pages}{054512} (\bibinfo{year}{2012}), \eprint{1110.6448}.

\bibitem[{\citenamefont{Bhattacharya
  et~al.}(2015{\natexlab{a}})\citenamefont{Bhattacharya, Cirigliano, Gupta,
  Lin, and Yoon}}]{Bhattacharya:2015esa}
\bibinfo{author}{\bibfnamefont{T.}~\bibnamefont{Bhattacharya}},
  \bibinfo{author}{\bibfnamefont{V.}~\bibnamefont{Cirigliano}},
  \bibinfo{author}{\bibfnamefont{R.}~\bibnamefont{Gupta}},
  \bibinfo{author}{\bibfnamefont{H.-W.} \bibnamefont{Lin}}, \bibnamefont{and}
  \bibinfo{author}{\bibfnamefont{B.}~\bibnamefont{Yoon}},
  \bibinfo{journal}{Phys. Rev. Lett.} \textbf{\bibinfo{volume}{115}},
  \bibinfo{pages}{212002} (\bibinfo{year}{2015}{\natexlab{a}}),
  \eprint{1506.04196}.

\bibitem[{\citenamefont{Follana et~al.}(2007)\citenamefont{Follana, Mason,
  Davies, Hornbostel, Lepage, Shigemitsu, Trottier, and Wong}}]{Follana:2006rc}
\bibinfo{author}{\bibfnamefont{E.}~\bibnamefont{Follana}},
  \bibinfo{author}{\bibfnamefont{Q.}~\bibnamefont{Mason}},
  \bibinfo{author}{\bibfnamefont{C.}~\bibnamefont{Davies}},
  \bibinfo{author}{\bibfnamefont{K.}~\bibnamefont{Hornbostel}},
  \bibinfo{author}{\bibfnamefont{G.~P.} \bibnamefont{Lepage}},
  \bibinfo{author}{\bibfnamefont{J.}~\bibnamefont{Shigemitsu}},
  \bibinfo{author}{\bibfnamefont{H.}~\bibnamefont{Trottier}}, \bibnamefont{and}
  \bibinfo{author}{\bibfnamefont{K.}~\bibnamefont{Wong}}
  (\bibinfo{collaboration}{HPQCD, UKQCD}), \bibinfo{journal}{Phys. Rev.}
  \textbf{\bibinfo{volume}{D75}}, \bibinfo{pages}{054502}
  (\bibinfo{year}{2007}), \eprint{hep-lat/0610092}.

\bibitem[{\citenamefont{Bazavov et~al.}(2013)}]{Bazavov:2012xda}
\bibinfo{author}{\bibfnamefont{A.}~\bibnamefont{Bazavov}} \bibnamefont{et~al.}
  (\bibinfo{collaboration}{MILC}), \bibinfo{journal}{Phys. Rev.}
  \textbf{\bibinfo{volume}{D87}}, \bibinfo{pages}{054505}
  (\bibinfo{year}{2013}), \eprint{1212.4768}.

\bibitem[{\citenamefont{Babich et~al.}(2010)\citenamefont{Babich, Brannick,
  Brower, Clark, Manteuffel, McCormick, Osborn, and Rebbi}}]{Babich:2010qb}
\bibinfo{author}{\bibfnamefont{R.}~\bibnamefont{Babich}},
  \bibinfo{author}{\bibfnamefont{J.}~\bibnamefont{Brannick}},
  \bibinfo{author}{\bibfnamefont{R.~C.} \bibnamefont{Brower}},
  \bibinfo{author}{\bibfnamefont{M.~A.} \bibnamefont{Clark}},
  \bibinfo{author}{\bibfnamefont{T.~A.} \bibnamefont{Manteuffel}},
  \bibinfo{author}{\bibfnamefont{S.~F.} \bibnamefont{McCormick}},
  \bibinfo{author}{\bibfnamefont{J.~C.} \bibnamefont{Osborn}},
  \bibnamefont{and} \bibinfo{author}{\bibfnamefont{C.}~\bibnamefont{Rebbi}},
  \bibinfo{journal}{Phys. Rev. Lett.} \textbf{\bibinfo{volume}{105}},
  \bibinfo{pages}{201602} (\bibinfo{year}{2010}), \eprint{1005.3043}.

\bibitem[{\citenamefont{Osborn et~al.}(2010)\citenamefont{Osborn, Babich,
  Brannick, Brower, Clark, Cohen, and Rebbi}}]{Osborn:2010mb}
\bibinfo{author}{\bibfnamefont{J.~C.} \bibnamefont{Osborn}},
  \bibinfo{author}{\bibfnamefont{R.}~\bibnamefont{Babich}},
  \bibinfo{author}{\bibfnamefont{J.}~\bibnamefont{Brannick}},
  \bibinfo{author}{\bibfnamefont{R.~C.} \bibnamefont{Brower}},
  \bibinfo{author}{\bibfnamefont{M.~A.} \bibnamefont{Clark}},
  \bibinfo{author}{\bibfnamefont{S.~D.} \bibnamefont{Cohen}}, \bibnamefont{and}
  \bibinfo{author}{\bibfnamefont{C.}~\bibnamefont{Rebbi}},
  \bibinfo{journal}{PoS} \textbf{\bibinfo{volume}{LATTICE2010}},
  \bibinfo{pages}{037} (\bibinfo{year}{2010}), \eprint{1011.2775}.

\bibitem[{\citenamefont{Edwards and Joo}(2005)}]{Edwards:2004sx}
\bibinfo{author}{\bibfnamefont{R.~G.} \bibnamefont{Edwards}} \bibnamefont{and}
  \bibinfo{author}{\bibfnamefont{B.}~\bibnamefont{Joo}}
  (\bibinfo{collaboration}{SciDAC, LHPC, UKQCD}), \bibinfo{journal}{Nucl. Phys.
  Proc. Suppl.} \textbf{\bibinfo{volume}{140}}, \bibinfo{pages}{832}
  (\bibinfo{year}{2005}), \bibinfo{note}{[,832(2004)]},
  \eprint{hep-lat/0409003}.

\bibitem[{\citenamefont{Yoon et~al.}(2016)}]{Yoon:2016dij}
\bibinfo{author}{\bibfnamefont{B.}~\bibnamefont{Yoon}} \bibnamefont{et~al.},
  \bibinfo{journal}{Phys. Rev.} \textbf{\bibinfo{volume}{D93}},
  \bibinfo{pages}{114506} (\bibinfo{year}{2016}), \eprint{1602.07737}.

\bibitem[{\citenamefont{Pedregosa et~al.}(2011)\citenamefont{Pedregosa,
  Varoquaux, Gramfort, Michel, Thirion, Grisel, Blondel, Prettenhofer, Weiss,
  Dubourg et~al.}}]{scikit-learn}
\bibinfo{author}{\bibfnamefont{F.}~\bibnamefont{Pedregosa}},
  \bibinfo{author}{\bibfnamefont{G.}~\bibnamefont{Varoquaux}},
  \bibinfo{author}{\bibfnamefont{A.}~\bibnamefont{Gramfort}},
  \bibinfo{author}{\bibfnamefont{V.}~\bibnamefont{Michel}},
  \bibinfo{author}{\bibfnamefont{B.}~\bibnamefont{Thirion}},
  \bibinfo{author}{\bibfnamefont{O.}~\bibnamefont{Grisel}},
  \bibinfo{author}{\bibfnamefont{M.}~\bibnamefont{Blondel}},
  \bibinfo{author}{\bibfnamefont{P.}~\bibnamefont{Prettenhofer}},
  \bibinfo{author}{\bibfnamefont{R.}~\bibnamefont{Weiss}},
  \bibinfo{author}{\bibfnamefont{V.}~\bibnamefont{Dubourg}},
  \bibnamefont{et~al.}, \bibinfo{journal}{Journal of Machine Learning Research}
  \textbf{\bibinfo{volume}{12}}, \bibinfo{pages}{2825} (\bibinfo{year}{2011}).

\bibitem[{\citenamefont{Breiman et~al.}(1984)\citenamefont{Breiman, Friedman,
  Stone, and Olshen}}]{breiman1984classification}
\bibinfo{author}{\bibfnamefont{L.}~\bibnamefont{Breiman}},
  \bibinfo{author}{\bibfnamefont{J.}~\bibnamefont{Friedman}},
  \bibinfo{author}{\bibfnamefont{C.}~\bibnamefont{Stone}}, \bibnamefont{and}
  \bibinfo{author}{\bibfnamefont{R.}~\bibnamefont{Olshen}},
  \emph{\bibinfo{title}{Classification and Regression Trees}}, The Wadsworth
  and Brooks-Cole statistics-probability series (\bibinfo{publisher}{Taylor \&
  Francis}, \bibinfo{year}{1984}), ISBN \bibinfo{isbn}{9780412048418},
  \urlprefix\url{https://books.google.com/books?id=JwQx-WOmSyQC}.

\bibitem[{\citenamefont{Friedman}(2000)}]{Friedman00greedyfunction}
\bibinfo{author}{\bibfnamefont{J.~H.} \bibnamefont{Friedman}},
  \bibinfo{journal}{Annals of Statistics} \textbf{\bibinfo{volume}{29}},
  \bibinfo{pages}{1189} (\bibinfo{year}{2000}).

\bibitem[{\citenamefont{Friedman}(2002)}]{Friedman:2002:SGB:635939.635941}
\bibinfo{author}{\bibfnamefont{J.~H.} \bibnamefont{Friedman}},
  \bibinfo{journal}{Comput. Stat. Data Anal.} \textbf{\bibinfo{volume}{38}},
  \bibinfo{pages}{367} (\bibinfo{year}{2002}), ISSN \bibinfo{issn}{0167-9473},
  \urlprefix\url{http://dx.doi.org/10.1016/S0167-9473(01)00065-2}.

\bibitem[{\citenamefont{Santosa and Symes}(1986)}]{Santosa:1986}
\bibinfo{author}{\bibfnamefont{F.}~\bibnamefont{Santosa}} \bibnamefont{and}
  \bibinfo{author}{\bibfnamefont{W.~W.} \bibnamefont{Symes}},
  \bibinfo{journal}{SIAM J. Sci. and Stat. Comput.}
  \textbf{\bibinfo{volume}{7(4)}}, \bibinfo{pages}{1307}
  (\bibinfo{year}{1986}).

\bibitem[{\citenamefont{Tibshirani}(1994)}]{Tibshirani94regressionshrinkage}
\bibinfo{author}{\bibfnamefont{R.}~\bibnamefont{Tibshirani}},
  \bibinfo{journal}{Journal of the Royal Statistical Society. Series B}
  \textbf{\bibinfo{volume}{58}}, \bibinfo{pages}{267} (\bibinfo{year}{1994}).

\bibitem[{\citenamefont{Hoerl and Kennard}(1970)}]{Hoerl1}
\bibinfo{author}{\bibfnamefont{A.~E.} \bibnamefont{Hoerl}} \bibnamefont{and}
  \bibinfo{author}{\bibfnamefont{R.~W.} \bibnamefont{Kennard}},
  \bibinfo{journal}{Technometrics} \textbf{\bibinfo{volume}{12}},
  \bibinfo{pages}{55} (\bibinfo{year}{1970}).

\bibitem[{\citenamefont{Bhattacharya et~al.}(2014)\citenamefont{Bhattacharya,
  Cohen, Gupta, Joseph, Lin, and Yoon}}]{Bhattacharya:2013ehc}
\bibinfo{author}{\bibfnamefont{T.}~\bibnamefont{Bhattacharya}},
  \bibinfo{author}{\bibfnamefont{S.~D.} \bibnamefont{Cohen}},
  \bibinfo{author}{\bibfnamefont{R.}~\bibnamefont{Gupta}},
  \bibinfo{author}{\bibfnamefont{A.}~\bibnamefont{Joseph}},
  \bibinfo{author}{\bibfnamefont{H.-W.} \bibnamefont{Lin}}, \bibnamefont{and}
  \bibinfo{author}{\bibfnamefont{B.}~\bibnamefont{Yoon}},
  \bibinfo{journal}{Phys. Rev.} \textbf{\bibinfo{volume}{D89}},
  \bibinfo{pages}{094502} (\bibinfo{year}{2014}), \eprint{1306.5435}.

\bibitem[{\citenamefont{Pospelov and Ritz}(2005)}]{Pospelov:2005pr}
\bibinfo{author}{\bibfnamefont{M.}~\bibnamefont{Pospelov}} \bibnamefont{and}
  \bibinfo{author}{\bibfnamefont{A.}~\bibnamefont{Ritz}},
  \bibinfo{journal}{Annals Phys.} \textbf{\bibinfo{volume}{318}},
  \bibinfo{pages}{119} (\bibinfo{year}{2005}), \eprint{hep-ph/0504231}.

\bibitem[{\citenamefont{Ramsey-Musolf and Su}(2008)}]{RamseyMusolf:2006vr}
\bibinfo{author}{\bibfnamefont{M.~J.} \bibnamefont{Ramsey-Musolf}}
  \bibnamefont{and} \bibinfo{author}{\bibfnamefont{S.}~\bibnamefont{Su}},
  \bibinfo{journal}{Phys. Rept.} \textbf{\bibinfo{volume}{456}},
  \bibinfo{pages}{1} (\bibinfo{year}{2008}), \eprint{hep-ph/0612057}.

\bibitem[{\citenamefont{Engel et~al.}(2013)\citenamefont{Engel, Ramsey-Musolf,
  and van Kolck}}]{Engel:2013lsa}
\bibinfo{author}{\bibfnamefont{J.}~\bibnamefont{Engel}},
  \bibinfo{author}{\bibfnamefont{M.~J.} \bibnamefont{Ramsey-Musolf}},
  \bibnamefont{and} \bibinfo{author}{\bibfnamefont{U.}~\bibnamefont{van
  Kolck}}, \bibinfo{journal}{Prog. Part. Nucl. Phys.}
  \textbf{\bibinfo{volume}{71}}, \bibinfo{pages}{21} (\bibinfo{year}{2013}),
  \eprint{1303.2371}.

\bibitem[{\citenamefont{Syritsyn et~al.}(2018)\citenamefont{Syritsyn, Izubuchi,
  and Ohki}}]{Syritsyn:2018mon}
\bibinfo{author}{\bibfnamefont{S.}~\bibnamefont{Syritsyn}},
  \bibinfo{author}{\bibfnamefont{T.}~\bibnamefont{Izubuchi}}, \bibnamefont{and}
  \bibinfo{author}{\bibfnamefont{H.}~\bibnamefont{Ohki}}, in
  \emph{\bibinfo{booktitle}{{13th Conference on the Intersections of Particle
  and Nuclear Physics (CIPANP 2018) Palm Springs, California, USA, May 29-June
  3, 2018}}} (\bibinfo{year}{2018}), \eprint{1810.03721}.

\bibitem[{\citenamefont{Kim et~al.}(2018)\citenamefont{Kim, Dragos, Shindler,
  Luu, and de~Vries}}]{Kim:2018rce}
\bibinfo{author}{\bibfnamefont{J.}~\bibnamefont{Kim}},
  \bibinfo{author}{\bibfnamefont{J.}~\bibnamefont{Dragos}},
  \bibinfo{author}{\bibfnamefont{A.}~\bibnamefont{Shindler}},
  \bibinfo{author}{\bibfnamefont{T.}~\bibnamefont{Luu}}, \bibnamefont{and}
  \bibinfo{author}{\bibfnamefont{J.}~\bibnamefont{de~Vries}}, in
  \emph{\bibinfo{booktitle}{{36th International Symposium on Lattice Field
  Theory (Lattice 2018) East Lansing, MI, United States, July 22-28, 2018}}}
  (\bibinfo{year}{2018}), \eprint{1810.10301}.

\bibitem[{\citenamefont{Bhattacharya et~al.}(2018)\citenamefont{Bhattacharya,
  Yoon, Gupta, and Cirigliano}}]{Bhattacharya:2018qat}
\bibinfo{author}{\bibfnamefont{T.}~\bibnamefont{Bhattacharya}},
  \bibinfo{author}{\bibfnamefont{B.}~\bibnamefont{Yoon}},
  \bibinfo{author}{\bibfnamefont{R.}~\bibnamefont{Gupta}}, \bibnamefont{and}
  \bibinfo{author}{\bibfnamefont{V.}~\bibnamefont{Cirigliano}}, in
  \emph{\bibinfo{booktitle}{{36th International Symposium on Lattice Field
  Theory (Lattice 2018) East Lansing, MI, United States, July 22-28, 2018}}}
  (\bibinfo{year}{2018}), \eprint{1812.06233}.

\bibitem[{\citenamefont{Bhattacharya
  et~al.}(2016{\natexlab{b}})\citenamefont{Bhattacharya, Cirigliano, Gupta,
  Mereghetti, and Yoon}}]{Bhattacharya:2016oqm}
\bibinfo{author}{\bibfnamefont{T.}~\bibnamefont{Bhattacharya}},
  \bibinfo{author}{\bibfnamefont{V.}~\bibnamefont{Cirigliano}},
  \bibinfo{author}{\bibfnamefont{R.}~\bibnamefont{Gupta}},
  \bibinfo{author}{\bibfnamefont{E.}~\bibnamefont{Mereghetti}},
  \bibnamefont{and} \bibinfo{author}{\bibfnamefont{B.}~\bibnamefont{Yoon}},
  \bibinfo{journal}{PoS} \textbf{\bibinfo{volume}{LATTICE2015}},
  \bibinfo{pages}{238} (\bibinfo{year}{2016}{\natexlab{b}}),
  \eprint{1601.02264}.

\bibitem[{\citenamefont{Bhattacharya
  et~al.}(2016{\natexlab{c}})\citenamefont{Bhattacharya, Cirigliano, Gupta, and
  Yoon}}]{Bhattacharya:2016rrc}
\bibinfo{author}{\bibfnamefont{T.}~\bibnamefont{Bhattacharya}},
  \bibinfo{author}{\bibfnamefont{V.}~\bibnamefont{Cirigliano}},
  \bibinfo{author}{\bibfnamefont{R.}~\bibnamefont{Gupta}}, \bibnamefont{and}
  \bibinfo{author}{\bibfnamefont{B.}~\bibnamefont{Yoon}},
  \bibinfo{journal}{PoS} \textbf{\bibinfo{volume}{LATTICE2016}},
  \bibinfo{pages}{225} (\bibinfo{year}{2016}{\natexlab{c}}),
  \eprint{1612.08438}.

\bibitem[{\citenamefont{Bhattacharya
  et~al.}(2015{\natexlab{b}})\citenamefont{Bhattacharya, Cirigliano, Gupta,
  Mereghetti, and Yoon}}]{Bhattacharya:2015rsa}
\bibinfo{author}{\bibfnamefont{T.}~\bibnamefont{Bhattacharya}},
  \bibinfo{author}{\bibfnamefont{V.}~\bibnamefont{Cirigliano}},
  \bibinfo{author}{\bibfnamefont{R.}~\bibnamefont{Gupta}},
  \bibinfo{author}{\bibfnamefont{E.}~\bibnamefont{Mereghetti}},
  \bibnamefont{and} \bibinfo{author}{\bibfnamefont{B.}~\bibnamefont{Yoon}},
  \bibinfo{journal}{Phys. Rev.} \textbf{\bibinfo{volume}{D92}},
  \bibinfo{pages}{114026} (\bibinfo{year}{2015}{\natexlab{b}}),
  \eprint{1502.07325}.

\bibitem[{\citenamefont{Shintani et~al.}(2016)\citenamefont{Shintani, Blum,
  Izubuchi, and Soni}}]{Shintani:2015vsx}
\bibinfo{author}{\bibfnamefont{E.}~\bibnamefont{Shintani}},
  \bibinfo{author}{\bibfnamefont{T.}~\bibnamefont{Blum}},
  \bibinfo{author}{\bibfnamefont{T.}~\bibnamefont{Izubuchi}}, \bibnamefont{and}
  \bibinfo{author}{\bibfnamefont{A.}~\bibnamefont{Soni}},
  \bibinfo{journal}{Phys. Rev.} \textbf{\bibinfo{volume}{D93}},
  \bibinfo{pages}{094503} (\bibinfo{year}{2016}), \eprint{1512.00566}.

\end{thebibliography}

\end{document}